\documentclass[11pt,twoside]{article}
\usepackage{asp2010,graphicx}

\resetcounters

\newcommand{\kms}{km s$^{-1}$}
\newcommand{\vk}{$v_{\rm k}$}
\newcommand{\vesc}{$v_{\rm esc}$}
\newcommand{\tmrg}{$t_{\rm mrg}$}

\newcommand{\fgas}{$f_{\rm gas}$}
\newcommand{\tagn}{$t_{\rm AGN}$}

\begin{document}

\title{Recoiling Black Holes in Merging Galaxies: Relationship to AGN Lifetimes, Starbursts, and the M-sigma Relation}
\author{Laura~Blecha$^1$, Thomas~J.~Cox$^2$, Abraham~Loeb$^1$, and Lars~Hernquist$^1$
\affil{$^1$Harvard University, Department of Astronomy, 60 Garden St., Cambridge, MA 02138}
\affil{$^2$Carnegie Observatories, 813 Santa Barbara Street, Pasadena, CA 91101}}

\begin{abstract}
Central supermassive black holes (SMBHs) are a ubiquitous feature of locally-observed galaxies, and ample evidence suggests that the growth of SMBHs and their host galaxies is closely linked. However, in the event of a merger, gravitational-wave (GW) recoil may displace a SMBH from its galactic center, or eject it entirely. To explore the consequences of this phenomenon, we use hydrodynamic simulations of gaseous galaxy mergers that include a range of BH recoil velocities. We have generated a suite of over 200 simulations with more than 60 merger models, enabling us to identify systematic trends in the behavior of recoiling BHs -- specifically (i) their dynamics, (ii) their observable signatures, and (iii) their effects on BH/galaxy co-evolution. (i) Recoiling BH trajectories depend heavily on the gas content of the host galaxy; maximal BH displacements from the center may vary by up to an order of magnitude between gas-rich and gas-poor mergers. In some cases, recoil trajectories also depend on the timing of the BH merger relative to the formation of the galaxy merger remnant. (ii) Recoiling BHs may be observable as offset active galactic nuclei (AGN) via either kinematic offsets ($v > 800$ km s$^{-1}$) or spatial offsets ($R > 1$ kpc) for lifetimes of about 1 - 100 Myr. In addition, recoil events affect the total AGN lifetime. GW recoil generally reduces the lifetimes of bright AGN, but may extend lower-luminosity AGN lifetimes. (iii) Rapidly-recoiling BHs may be up to about 5 times less massive than their stationary counterparts. These mass deficits lower the normalization of the M - $\sigma$ relation and contribute to both intrinsic and overall scatter. Furthermore, recoil events displace AGN feedback from the galactic center, which enhances central star formation rates. This results in longer starburst phases and higher central stellar densities in merger remnants.
\end{abstract}

\section{Introduction}
Recent results from numerical relativity simulations indicate that ``central" supermassive black holes (SMBHs) may in fact spend substantial time in motion and offset from the galactic center, owing to the large gravitational-wave (GW) recoil kicks that may accompany BH mergers \citep[e.g.,][]{campan07a, baker08, lousto10b}. Recoiling BHs may be observable as offset AGN \cite[][]{madqua04, loeb07, bleloe08, blecha11,guedes11,sijack11}. Several such candidates have been discovered, though none have yet been confirmed \citep{komoss08, shield09b,civano10,batche10,jonker10}. There may also be indirect consequences of GW recoil for BH - galaxy co-evolution \citep{volont07,blecha11,sijack11}. 

We conduct a parameter study of recoiling BHs, using hydrodynamic simulations of galaxy mergers. Our simulations are described in \S~\ref{sec:sims}. We discuss recoiling BH dynamics in \S~\ref{ssec:traj}. In \S~\ref{ssec:agn}, we examine the offset lifetimes of recoiling AGN, and in \S~\ref{ssec:coev}, we examine the effects on host galaxies. Our conclusions are summarized in \S~\ref{sec:conc}.

\section{Galaxy Merger Simulations}
\label{sec:sims}

We simulate galaxy mergers using the smoothed particle hydrodynamics (SPH) code {\footnotesize GADGET-3} \citep{spring05a,spring05b}. We add an arbitrary kick velocity~\vk~to the remnant BH at the time of BH merger (\tmrg), which we generally scale to the central escape speed, \vesc(\tmrg). The BHs are allowed to accrete both from ambient gas via the Bondi-Hoyle formula and from an ejected disk of gas with a time-dependent accretion rate. These prescriptions are described in detail in \citet{blecha11}.

We have used 62 different galaxy merger models in which we vary the galaxy mass ratio ($q$), the total galaxy mass, the gas fraction (\fgas), and the orbital configuration. For each model, we simulate both a merger with no recoil kick and a merger with~\vk/\vesc~$ = 0.9$.  For a subset of these models, we also simulate intermediate values of~\vk. We refer the reader to~\citet{blecha11} for the full details of our initial conditions.

\section{Results}

\subsection{Recoiling BH Trajectories}
\label{ssec:traj}

\begin{figure}
\center{\includegraphics[width=0.495\textwidth]{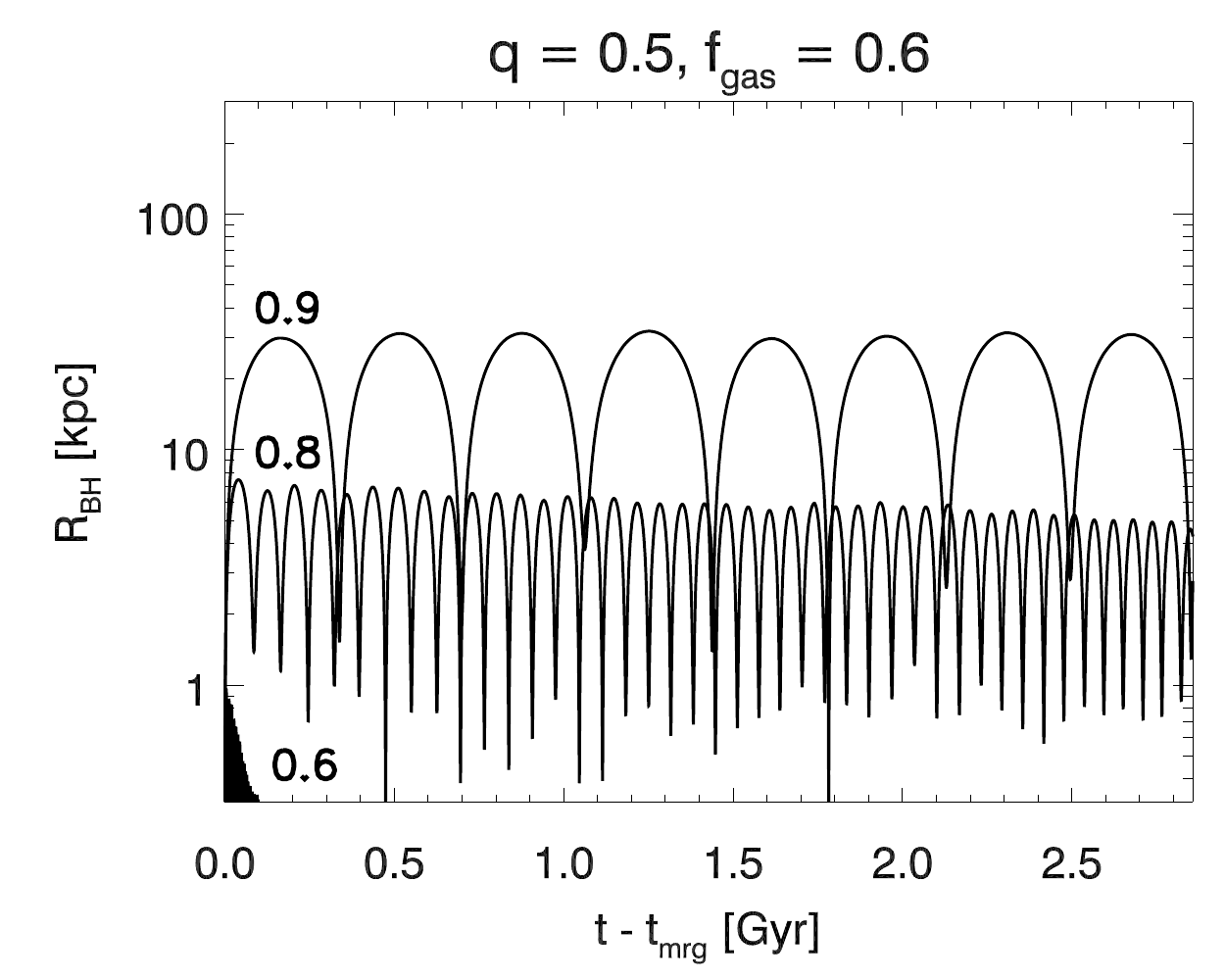}
\includegraphics[width=0.495\textwidth]{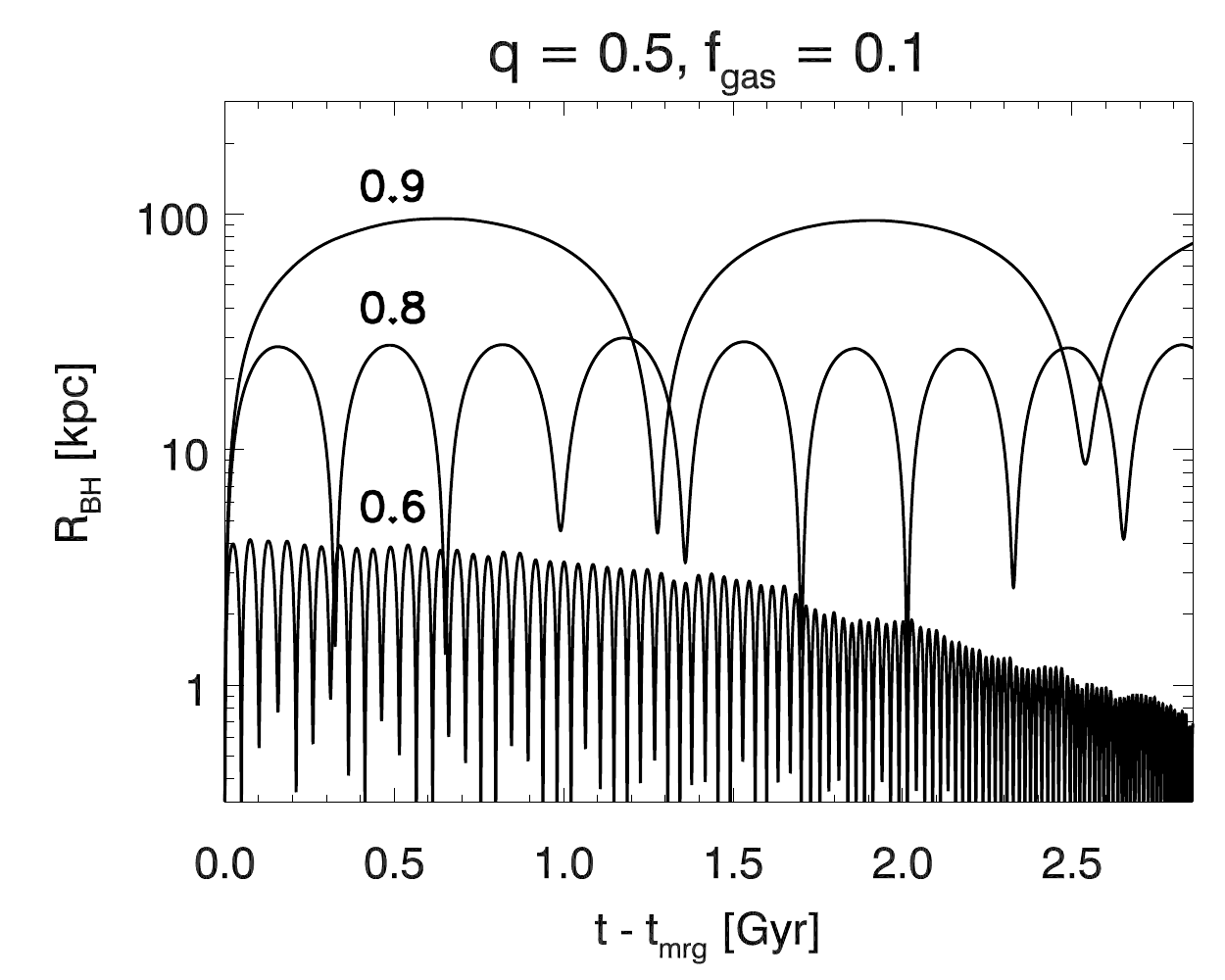}}
\caption[]{Recoil trajectories in gas-rich (left panel, 60\% gas initially) and gas-poor (right panel, 10\% gas) mergers. In each case, the recoiling BH separation from the galactic center is plotted versus time for varying~\vk/\vesc~within a single galaxy model. \vk/\vesc\ $=$ 0.6, 0.8, \& 0.9 are shown, as indicated on the plots.}
\label{fig:2traj}
\end{figure}

Fig.~\ref{fig:2traj} demonstrates that the gas content of galaxies greatly influences recoiling BH trajectories. In the left panel (right panel), trajectories with~\vk~$= 0.6,\, 0.8,\, \&\, 0.9$~\vesc~are shown for a merger in which the progenitor galaxies had 60\% (10\%) gas initially. In both cases, BHs kicked with \vk\ $= 0.9$ \vesc\ are still on large orbits by the end of the simulation. However, at more moderate kick speeds, the BH has a much shorter wandering time in the gas-rich remnant than in the gas-poor remnant. For  \vk\ $= 0.6$ \vesc, the BH in the gas-rich remnant is confined to the central kpc.

\subsection{AGN Lifetimes}
\label{ssec:agn}

\begin{figure}
\center{\includegraphics[width=0.83\textwidth]{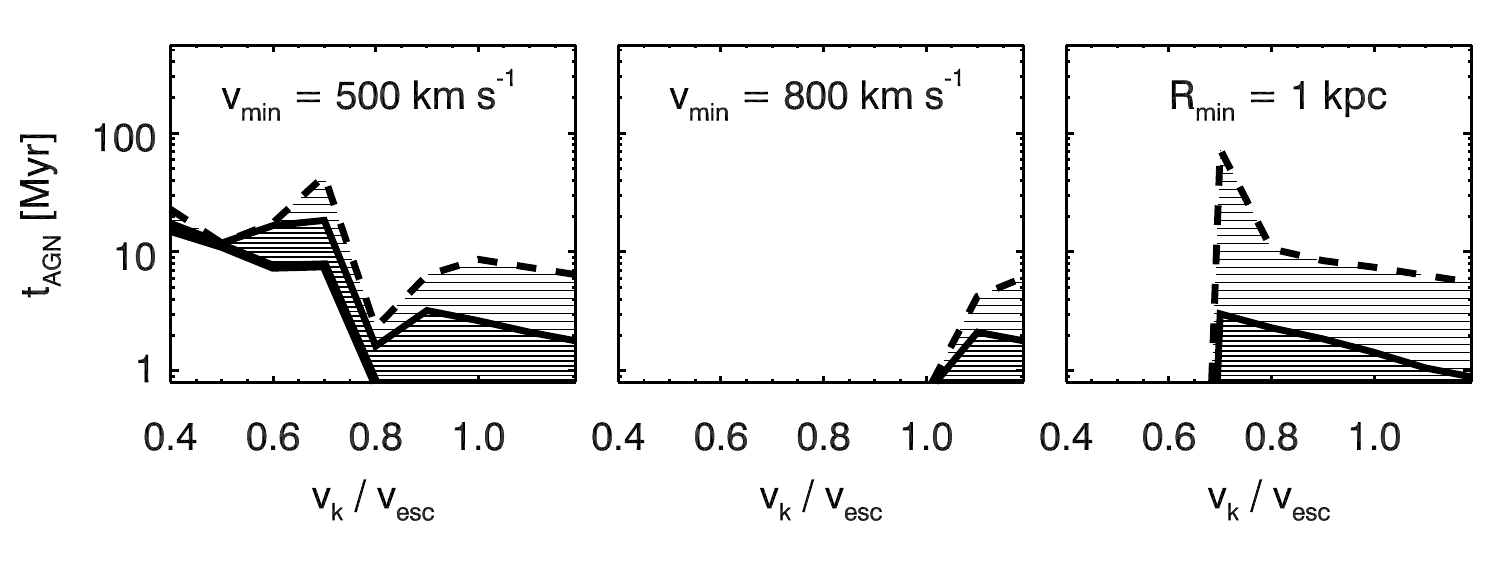}}
\caption[]{Left panel: Kinematically-offset AGN lifetimes for varying \vk/\vesc\, with $v_{\rm min} = 500$ \kms, for the 3 AGN definitions given in the text. Middle panel: same for $v_{\rm min} = 800$ \kms. Right panel: spatially-offset AGN lifetimes for $R_{\rm min} = 1$ kpc. \label{fig:tagn_vmin}}
\end{figure}

We calculate the observable lifetimes of recoiling BHs as spatially- or kinematically-offset AGN (\tagn). We use 3 different definitions of an AGN: $L_{\rm bol} > (10\%\, L_{\rm Edd}$, $3\%\, L_{\rm Edd}$, $3.3\times10^9$ L$_{\odot}$).  $L_{\rm bol}$ \& $L_{\rm Edd}$ are the bolometric and Eddington luminosities. For kinematically (spatially) offset AGN, we require a BH velocity offset $v_{\rm BH} > v_{\rm min}$ ($R_{\rm BH} > R_{\rm min}$) from the stellar center of mass. Fig.~\ref{fig:tagn_vmin} shows the resulting offset AGN lifetimes. At moderate kick speeds, a kinematically-offset AGN may occur when the BH passes through the gas-rich galactic center. At higher kick speeds with longer orbits, the BH appears as a kinematically offset AGN only immediately after the kick. Spatially-offset AGN may have longer lifetimes, but generally at lower luminosities.

\subsection{Co-evolution of BHs and Galaxies}
\label{ssec:coev}

\begin{figure}
\center{\includegraphics[width=0.5\textwidth]{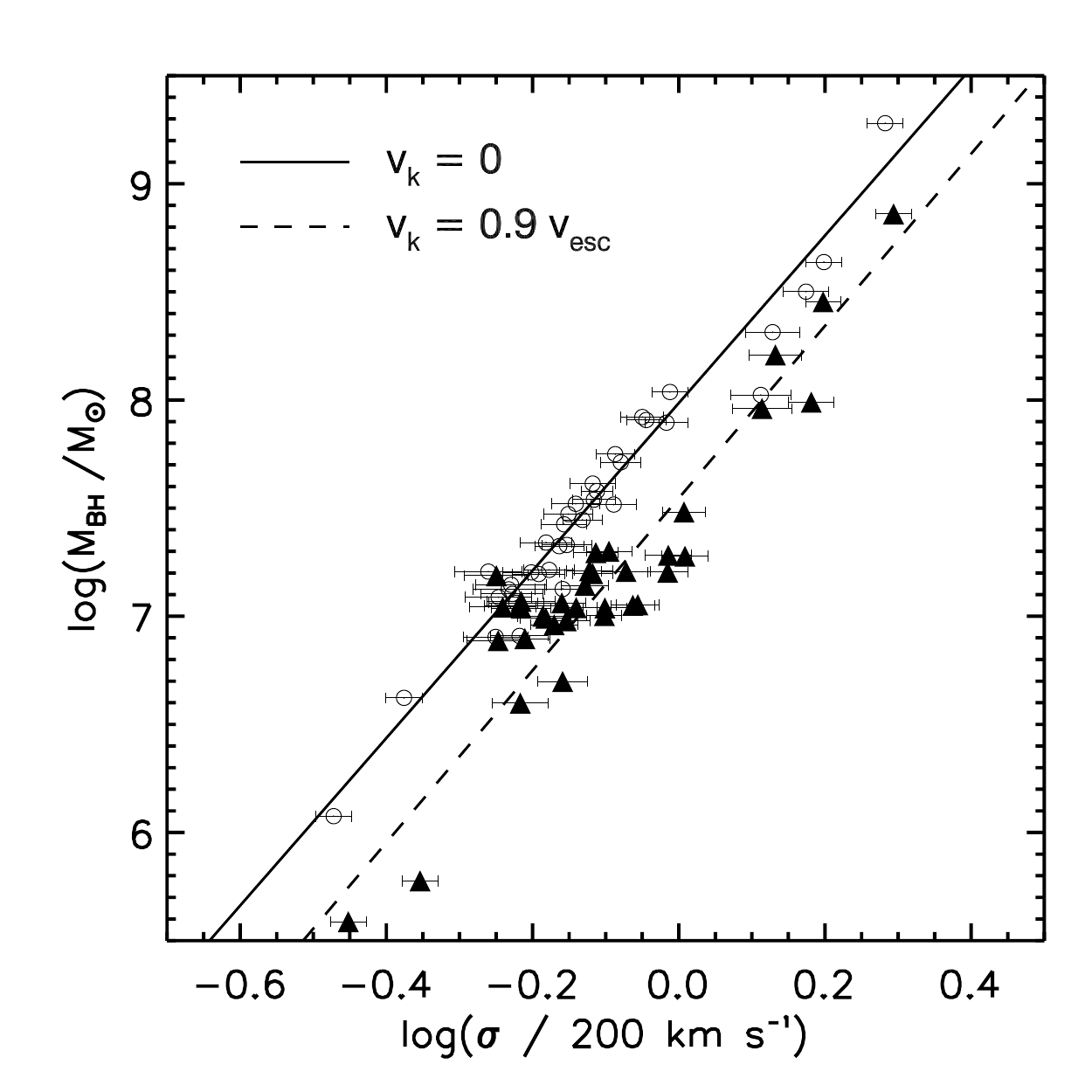}
\includegraphics[width=0.49\textwidth]{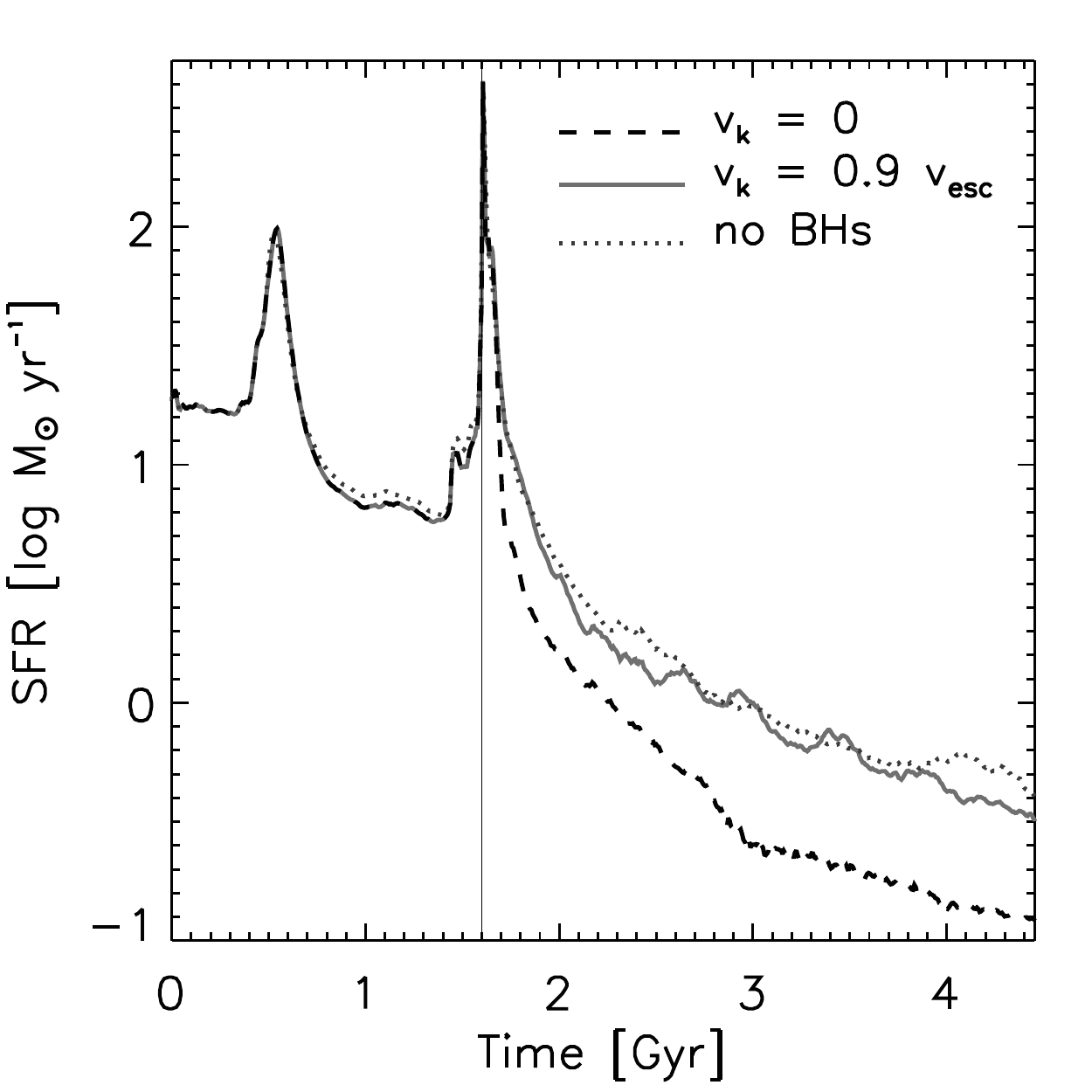}}
\caption[]{Left: $M_{\rm BH}-\sigma_*$ relation for simulations with no recoil kicks (open circles) and with~\vk/\vesc~$=0.9$ (filled triangles).  $M_{\rm BH}$ is the final BH mass and $\sigma_*$ is the line-of-sight-averaged stellar velocity dispersion; error bars show the range of $\sigma_*$.  The solid \& dashed lines are least-squares fits to the data. Right: Total SFR vs. time for a gas-rich merger.  The simulations shown have~\vk~$= 0$ (dashed line),~\vk/\vesc~$= 0.9$ (solid line), and no BHs (dotted line). The vertical line marks the time of BH merger. \label{fig:coeval}}
\end{figure}

Recoiling BHs have smaller final masses than stationary BHs, in some cases by up to a factor of $\sim 5$. In the left panel of Fig.~\ref{fig:coeval}, we compare the $M_{\rm BH} - \sigma_*$ relation resulting from our sets of simulations with no recoil and with large recoil (\vk/\vesc~$= 0.9$). These data do not reproduce the observed $M_{\rm BH} - \sigma_*$ relation~\cite[e.g.,][]{tremai02}, as we have varied parameters systematically, but the relative differences between the simulated correlations are informative. GW recoil lowers the offset of the $M_{\rm BH} - \sigma_*$ relation and increases intrinsic scatter.  In a cosmological framework, scatter would increase even further as some recoiling BHs were replaced by new BHs via subsequent mergers.

GW recoil can also affect the central structure of galaxy merger remnants. The right panel of Fig.~\ref{fig:coeval} compares the star formation rates (SFRs) for simulations of a gas-rich merger with~\vk/\vesc~$= 0$ and 0.9, and with no BHs. The simulation with large recoil has a higher post-merger SFR than the no-recoil simulation, similar to the case with no BHs. This is a direct result of the displacement of AGN feedback via GW recoil.

\section{Conclusions}
\label{sec:conc}
We find that recoiling BHs may appear as offset AGN for $\sim 1 - 100$ Myr, for a wide range in kick speeds. Their trajectories are greatly influenced by the environment of a gaseous merger remnant. Moreover, recoiling BHs themselves may influence the central regions of the merger remnant and contribute to scatter in the BH - bulge relations.

\bibliography{Blecha_L_bib}

\begin{thebibliography}{}
\expandafter\ifx\csname natexlab\endcsname\relax\def\natexlab#1{#1}\fi
\expandafter\ifx\csname url\endcsname\relax
  \def\url#1{\texttt{#1}}\fi
\expandafter\ifx\csname urlprefix\endcsname\relax\def\urlprefix{URL }\fi
\providecommand{\eprint}[2][]{\url{#2}}

\bibitem[{{Baker} et~al.(2008){Baker}, {Boggs}, {Centrella}, {Kelly},
  {McWilliams}, {Miller}, \& {van Meter}}]{baker08}
{Baker}, J.~G., {Boggs}, W.~D., {Centrella}, J., {Kelly}, B.~J., {McWilliams},
  S.~T., {Miller}, M.~C., \& {van Meter}, J.~R. 2008, \apjl, 682, L29.
  \eprint{0802.0416}

\bibitem[{{Batcheldor} et~al.(2010){Batcheldor}, {Robinson}, {Axon}, {Perlman},
  \& {Merritt}}]{batche10}
{Batcheldor}, D., {Robinson}, A., {Axon}, D.~J., {Perlman}, E.~S., \&
  {Merritt}, D. 2010, \apjl, 717, L6. \eprint{1005.2173}

\bibitem[{{Blecha} et~al.(2011){Blecha}, {Cox}, {Loeb}, \&
  {Hernquist}}]{blecha11}
{Blecha}, L., {Cox}, T.~J., {Loeb}, A., \& {Hernquist}, L. 2011, \mnras, 412,
  2154. \eprint{1009.4940}

\bibitem[{{Blecha} \& {Loeb}(2008)}]{bleloe08}
{Blecha}, L., \& {Loeb}, A. 2008, \mnras, 390, 1311. \eprint{0805.1420}

\bibitem[{{Campanelli} et~al.(2007){Campanelli}, {Lousto}, {Zlochower}, \&
  {Merritt}}]{campan07a}
{Campanelli}, M., {Lousto}, C., {Zlochower}, Y., \& {Merritt}, D. 2007, \apjl,
  659, L5. \eprint{arXiv:gr-qc/0701164}

\bibitem[{{Civano}(2010)}]{civano10}
{Civano}, F.~{\em et al.}. 2010, \apj, 717, 209. \eprint{1003.0020}

\bibitem[{{Guedes} et~al.(2011){Guedes}, {Madau}, {Mayer}, \&
  {Callegari}}]{guedes11}
{Guedes}, J., {Madau}, P., {Mayer}, L., \& {Callegari}, S. 2011, \apj, 729,
  125. \eprint{1008.2032}

\bibitem[{{Jonker} et~al.(2010){Jonker}, {Torres}, {Fabian}, {Heida},
  {Miniutti}, \& {Pooley}}]{jonker10}
{Jonker}, P.~G., {Torres}, M.~A.~P., {Fabian}, A.~C., {Heida}, M., {Miniutti},
  G., \& {Pooley}, D. 2010, \mnras, 407, 645. \eprint{1004.5379}

\bibitem[{{Komossa} et~al.(2008){Komossa}, {Zhou}, \& {Lu}}]{komoss08}
{Komossa}, S., {Zhou}, H., \& {Lu}, H. 2008, \apjl, 678, L81.
  \eprint{arXiv:0804.4585}

\bibitem[{{Loeb}(2007)}]{loeb07}
{Loeb}, A. 2007, Physical Review Letters, 99, 041103.
  \eprint{arXiv:astro-ph/0703722}

\bibitem[{{Lousto} et~al.(2010){Lousto}, {Campanelli}, {Zlochower}, \&
  {Nakano}}]{lousto10b}
{Lousto}, C.~O., {Campanelli}, M., {Zlochower}, Y., \& {Nakano}, H. 2010,
  Classical and Quantum Gravity, 27, 114006. \eprint{0904.3541}

\bibitem[{{Madau} \& {Quataert}(2004)}]{madqua04}
{Madau}, P., \& {Quataert}, E. 2004, \apjl, 606, L17.
  \eprint{arXiv:astro-ph/0403295}

\bibitem[{{Shields}(2009)}]{shield09b}
{Shields}, G.~A.~{\em et al.}. 2009, \apj, 707, 936. \eprint{0907.3470}

\bibitem[{{Sijacki} et~al.(2011){Sijacki}, {Springel}, \&
  {Haehnelt}}]{sijack11}
{Sijacki}, D., {Springel}, V., \& {Haehnelt}, M.~G. 2011, \mnras, 414, 3656.
  \eprint{1008.3313}

\bibitem[{{Springel}(2005)}]{spring05a}
{Springel}, V. 2005, \mnras, 364, 1105. \eprint{arXiv:astro-ph/0505010}

\bibitem[{{Springel} et~al.(2005){Springel}, {Di Matteo}, \&
  {Hernquist}}]{spring05b}
{Springel}, V., {Di Matteo}, T., \& {Hernquist}, L. 2005, \mnras, 361, 776.
  \eprint{arXiv:astro-ph/0411108}

\bibitem[{{Tremaine}(2002)}]{tremai02}
{Tremaine}, S.~{\em et al.}. 2002, \apj, 574, 740.
  \eprint{arXiv:astro-ph/0203468}

\bibitem[{{Volonteri}(2007)}]{volont07}
{Volonteri}, M. 2007, \apjl, 663, L5. \eprint{arXiv:astro-ph/0703180}

\end{thebibliography}

\end{document}